\documentclass[aps,prl,twocolumn,showpacs,letterpaper,superscriptaddress]{revtex4-1}
\usepackage{graphicx}
\usepackage{multirow}
\usepackage{amsmath}
\usepackage{color}
\usepackage{hyperref}
\hypersetup{
    colorlinks=true,
    linkcolor=blue,
    filecolor=blue,
    urlcolor=blue,
    citecolor=blue,
}
\usepackage[normalem]{ulem}

\begin{document}

\title{Anomalous excitons and screenings unveiling strong electronic correlations in SrTi$_{1-x}$Nb$_x$O$_3$, 0$\leq$\textit{x}$\leq$0.005}
\affiliation{Singapore Synchrotron Light Source, National University of Singapore, 5 Research Link, Singapore 117603}
\affiliation{NUSNNI-NanoCore, National University of Singapore, Singapore 117576}
\affiliation{Department of Physics, Faculty of Science, National University of Singapore, Singapore 117542}
\affiliation{Centre for Advanced 2D Materials and Graphene Reseach Centre, National University of Singapore, Singapore, 117546}
\affiliation{Laboratoire des Solides Irradi\'es, \'Ecole Polytechnique, CNRS, CEA-DSM, F-91128 Palaiseau, France}
\affiliation{Institut N\'eel, CNRS, BP 166, F-38042 Grenoble, France}
\affiliation{Graduate School of Frontier Biosciences and Department of Physics, Osaka University, Osaka 565-0871, Japan}

\author{Pranjal Kumar Gogoi}
\affiliation{Singapore Synchrotron Light Source, National University of Singapore, 5 Research Link, Singapore 117603}
\affiliation{NUSNNI-NanoCore, National University of Singapore, Singapore 117576}
\affiliation{Department of Physics, Faculty of Science, National University of Singapore, Singapore 117542}

\author{Lorenzo Sponza}
\affiliation{Laboratoire des Solides Irradi\'es, \'Ecole Polytechnique, CNRS, CEA-DSM, F-91128 Palaiseau, France}

\author{Daniel Schmidt}
\affiliation{Singapore Synchrotron Light Source, National University of Singapore, 5 Research Link, Singapore 117603}

\author{Teguh Citra Asmara}
\affiliation{Singapore Synchrotron Light Source, National University of Singapore, 5 Research Link, Singapore 117603}
\affiliation{NUSNNI-NanoCore, National University of Singapore, Singapore 117576}
\affiliation{Department of Physics, Faculty of Science, National University of Singapore, Singapore 117542}

\author{Caozheng Diao}
\affiliation{Singapore Synchrotron Light Source, National University of Singapore, 5 Research Link, Singapore 117603}

\author{Jason C. W. Lim}
\affiliation{Singapore Synchrotron Light Source, National University of Singapore, 5 Research Link, Singapore 117603}
\affiliation{NUSNNI-NanoCore, National University of Singapore, Singapore 117576}

\author{Sock Mui Poh}
\affiliation{Singapore Synchrotron Light Source, National University of Singapore, 5 Research Link, Singapore 117603}

\author{Shin-ichi Kimura}
\affiliation{Graduate School of Frontier Biosciences and Department of Physics, Osaka University, Osaka 565-0871, Japan}

\author{Paolo E. Trevisanutto}
\affiliation{Singapore Synchrotron Light Source, National University of Singapore, 5 Research Link, Singapore 117603}
\affiliation{Department of Physics, Faculty of Science, National University of Singapore, Singapore 117542}
\affiliation{Centre for Advanced 2D Materials and Graphene Reseach Centre, National University of Singapore, Singapore, 117546}

\author{Valerio Olevano}
\email{valerio.olevano@grenoble.cnrs.fr}
\affiliation{Institut N\'eel, CNRS, BP 166, F-38042 Grenoble, France}

\author{Andrivo Rusydi}
\email{phyandri@nus.edu.sg}
\affiliation{Singapore Synchrotron Light Source, National University of Singapore, 5 Research Link, Singapore 117603}
\affiliation{NUSNNI-NanoCore, National University of Singapore, Singapore 117576}
\affiliation{Department of Physics, Faculty of Science, National University of Singapore, Singapore 117542}

\date{\today}
%
%abstract
%
\begin{abstract}
Electron-electron (e-e) and electron-hole (e-h) interactions are often associated with many exotic phenomena in correlated electron systems. Here, we report an observation of anomalous excitons at 3.75 , 4.67  and 6.11 eV at 4.2 K in \textit{bulk}-SrTiO$_3$. Fully supported by \textit{ab initio} GW Bethe-Salpeter equation calculations, these excitons are due to surprisingly strong e-h and e-e interactions with different characters: 4.67  and 6.11 eV are resonant excitons and 3.75 eV is a bound Wannier-\textit{like} exciton with an  unexpectedly higher level of delocalization. Measurements and calculations on SrTi$_{1-x}$Nb$_x$O$_3$ for 0.0001$\leq$\textit{x}$\leq$0.005 further show that energy and spectral-weight of the excitonic peaks vary as a function of electron doping (\textit{x}) and temperature, which are attributed to screening effects. Our results show the importance of e-h and e-e interactions yielding to anomalous excitons and thus bring out a new fundamental perspective in SrTiO$_3$.
\end{abstract}

%\pacs{71.10.Ca,71.15.Mb,71.45.Gm}

\maketitle

\begin{center}
{\bf I. INTRODUCTION}
\end{center}

Strong electronic correlations in transition metal oxides are known to drive rich phenomena such as superconductivity, colossal magnetoresistance and metal-insulator transitions \cite{Basov}. An example of a model transition metal oxide is the perovskite-type SrTiO$_3$ (STO). Indeed, STO in various bulk forms and recently its heterostructures with other oxides, have shown exotic phenomena such as superconductivity \cite{Schooley, Reyren,  Kozuka, Li, Bert}, magnetisms \cite{ Li, Bert, Brinkman, Ariando, Rice}, metal-insulator transitions and two-dimensional electron gas \cite{Ohtomo, Thiel, Santander-Syro, Meevasana, Asmara}. However the role of electronic correlations, particularly electron-hole (e-h) and electron-electron (e-e) interactions, remains unclear. In the past, it was assumed that the role of e-h and e-e interactions in STO was negligible, if any, and thus most calculations were done without implicitly incorporating such many-body effects \cite{ Kahn, Mattheiss, Ahuja, Gupta}. Recently, a theoretical study by solving the ab initio Bethe-Salpeter equation (BSE) within the GW approximation has shown that if such many-body effects are present in STO, this might yield to the so-called resonant excitonic effects and a prediction of a resonant exciton at $\sim$6.4 eV, which then should all be manifested in its complex dielectric function within a broad energy range \cite{Sponza}. If this is true, it makes STO even more interesting and important to study. Until we solve this fundamental problem, we may not be able to understand, for instance, the origin of such exotic phenomena as well as to resolve discrepancies in previous optical spectroscopic studies particularly near absorption edge characteristics in STO \cite{Cardona, Cohen, Capizzi, Bauerle, Servoin, Jellison, Zollner, Benthem}. An ideal tool to study resonant excitonic effects is spectroscopic ellipsometry (SE) because it probes neutral excitations, e.g. e-h pairs and excitons, providing important information in particular on two-particle correlations. Indeed, SE has been used and was successful to unravel the presence of resonant excitonic effects in graphene \cite{Mak, Santoso} as theoretically predicted \cite{Yang, Trevisanutto}.

In this work, systematic temperature-dependent SE measurements supported  by theoretical \textit{ab initio} BSE calculations on top of a GW electronic structure (GW-BSE) are used to unravel the presence of anomalous excitonic effects and thus to identify the role of e-h and e-e interactions in SrTi$_{1-x}$Nb$_x$O$_3$ for 0$\leq$x$\leq$0.005. Measurements and calculations on SrTi$_{1-x}$Nb$_x$O$_3$ are performed to determine effects of charge carrier screening on the optical spectra.

%%%%%%%%%%%%%%%%%%%%%%%%%%%%%%%%%%%%%%%%%%%%%%%%%%%%
\begin{figure}
\includegraphics[width=\columnwidth]{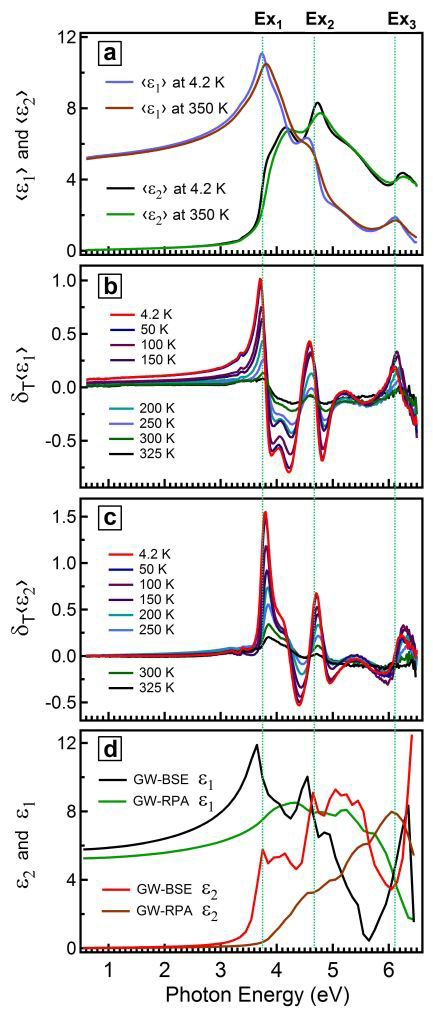}
\caption{\label{fig1} (Color online)
Experimental critical points and comparison with theory for STO. (a) $\langle\varepsilon_1\rangle$  and $\langle\varepsilon_2\rangle$  for the lowest (4.2 K) and highest (350 K) measured temperatures. (b), (c) Temperature dependent differences $\delta_T\langle\varepsilon_i\rangle = \langle\varepsilon_i\rangle(T) - \langle\varepsilon_i\rangle(T =350 K), i = 1, 2$. (d) Plots of $\varepsilon_1$ and $\varepsilon_2$ calculated using GW-BSE and GW-RPA (all graphs are blue-shifted by 47 meV to match experimental data). The green vertical dashed line are  determined using second derivative analysis.}
\end{figure}
%%%%%%%%%%%%%%%%%%%%%%%%%%%%%%%%%%%%%%%%%%%%%%%%%%%%

\begin{center}
{\bf II. SAMPLE PREPARATIONS, EXPERIMENTAL METHODS AND THEORETICAL CALCULATIONS}
\end{center}

SrTiO$_3$ (100) and  SrTi$_{1-x}$Nb$_x$O$_3$ (100), $x =$ 0.0001, 0.0005, 0.001 and 0.005  samples procured from Crystec are used for the measurements. All samples are single side polished of size 10 mm$ \times $ 10 mm $\times$ 0.5  mm. Atomic force microscopy (AFM) measurements show that the rms roughness for all the samples are less than 5 \AA.

Spectroscopic ellipsometry (SE) is a technique that measures the polarization state change of light upon reflection from (or transmission through) a sample. The measured ellipsometric angles $\Psi$ and $\Delta$ are related to the material properties of the sample by the complex reflectance ratio
\begin{equation} \label{ellips}
\varrho = {r_p}/{r_s} = \tan\Psi e^{i\Delta}
\end{equation}

where $r_p$   and $r_s$   are the Fresnel reflection coefficients for $p$ and $s$ polarized light, respectively \cite{Fujiwara}.

Spectroscopic ellipsometry is performed with a commercial rotating analyser ellipsometer (J.A. Woollam, Inc. V-VASE) covering the energy range of 0.6 -- 6.5 eV. Low temperature measurements are performed within a cryostat (Janis) and at base pressures in the $10^{-9}$ torr regime. Liquid helium and liquid nitrogen are used in the open cycle cryostat. The angle of incidence for all cryostat measurements is $70^\circ$. Major focus is put on the pseudo-dielectric function (PDF) as it can be directly obtained from the measured ellipsometric quantities ($\Psi$, $\Delta$) and it is identical to the complex dielectric function  $(\varepsilon_1, \varepsilon_2)$ for an isotropic bulk sample  with a perfecly smooth top surface \cite {Fujiwara}. With an optical model assuming a perfectly flat substrate having infinite thickness, the PDF $(\langle\varepsilon_1\rangle, \langle\varepsilon_2\rangle)$  can be obtained from the measured $(\Psi, \Delta)$  directly. In the experimental set up SrTiO$_3$ (STO) and SrTi$_{1-x}$Nb$_x$O$_3$ samples are measured in air and vacuum and hence the optical model is taken as an $air/sample$ or $vacuum/sample$ interface. In this case the PDF can be calculated using the following expression \cite{Fujiwara}:
\begin{equation}
\langle\varepsilon\rangle = {\sin}^2{\theta_i}\biggl[1 + {\tan}^2{\theta_i}{\biggl(\frac{1-\varrho}{1 + \varrho}\biggr)}^2\biggr]
\end{equation}
Where  $\theta_i$  is the incident angle and $\varrho$  is related to the measured $(\Psi, \Delta)$    by the expression (\ref{ellips}).

The GW-BSE calculations are performed for \textit{x} = 0 and 0.005 emphasizing the role of e-h and e-e interactions, particularly on near edge spectra, as well as screening effects in electron doping case. Note that neither a Kramers-Kronig transformation nor optical modelling is required here to further ensure that the observed optical phenomena are truly intrinsic property of the system under study.

%%%%%%%%%%%%%%%%%%%%%%%%%%%%%%%%%%%%%%%%%%%%%%%%%%%%
\begin{figure}
\centering
\includegraphics[width =\columnwidth] {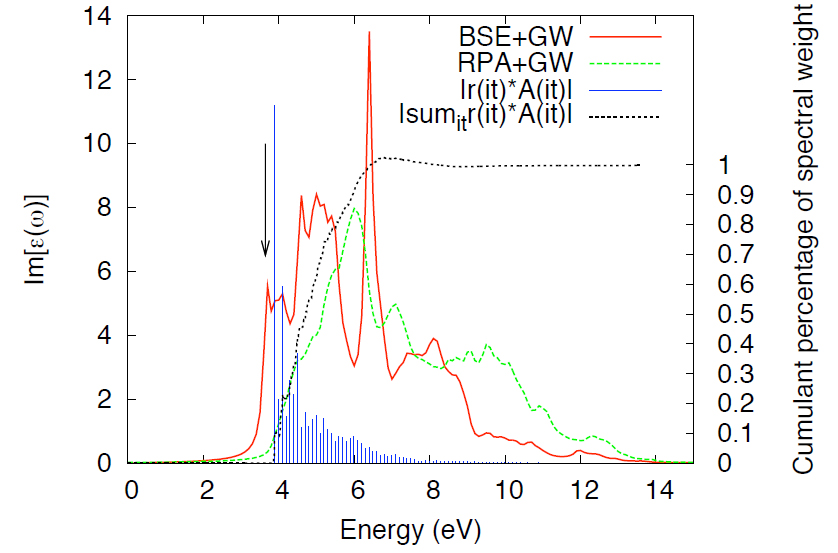}\\[5pt]
\caption {\label{SupplFig4} (Color online) GW-BSE analysis of the peak Ex$_1$. Peak Ex$_1$: $ E_\lambda= 3.644 $ eV. GW-BSE spectrum and IP-spectrum are represented by the red and green lines respectively. The histogram of $|{\tilde{\rho}}_tA_\lambda^t |$  is presented in blue. The cumulant weight function is reported in dotted black. A  black arrow indicates the energy $E_\lambda= 3.644 $eV.}
\end{figure}
%%%%%%%%%%%%%%%%%%%%%%%%%%%%%%%%%%%%%%%%%%%%%%%%%%%%

To solve the static BSE problem, it can be convenient to adopt a basis of independent-particle (IP) transitions \cite{onida}. In this basis the excitonic eigenvalue problem reads
\begin{equation}
\sum_{t^\prime} H_{t,t^\prime}A_{\lambda}^{t^\prime} = E_\lambda A_\lambda^t
\end{equation}
where $t$  and $t^\prime$   label the IP-transitions, $A_\lambda$  is the eigenvector relative to the eigenvalue $E_\lambda$ . The BSE absorption spectrum is determined calculating the macroscopic dielectric function:
\begin{equation}\label{eqn4}
\epsilon_M(\omega) = 1 - \lim_{{\bf q}\to {\bf 0}} v_0({\bf q})\sum_{\lambda}\frac{{\left|\sum_{t}\tilde{\rho_t}(q)A_\lambda^t\right|}^2}{E_\lambda - \omega - i\eta}
\end{equation}

where $v_0({\bf q})$  is the Coulomb interaction, $\eta$  is a positive infinitesimal quantity and,
\begin{equation}
{\tilde{\rho}}_{vc}(q)= \langle v|e^{{-i}\bf{q}\cdot\bf{r}}|c\rangle
\end{equation}
where  $c$ and $v$ are conduction and valence states respectively. The peaks are found for exciton energies $E_\lambda$  (the zeroes of the denominator in Equation (\ref{eqn4})).The weight of the excitation at $E_\lambda$ is given by the squared modulus of a sum over IP-transitions (the numerator of in Equation (\ref{eqn4})), namely
\begin{equation}
S_{\lambda}^{\infty} = {\bigg| \sum_{t}S_{\lambda}^t(\bf{q})\bigg|}^2\quad \text {with} \quad S_{\lambda}^t({\bf q}) = {\tilde{\rho}}_t({\bf{q}})A_\lambda^t
\end{equation}
where $S_{\lambda}^t ({\bf q})$   are the IP-weights and the sum is made over all transitions of the system (converged spectrum), which is stressed by the $\infty$ symbol.

For a given peak at energy $E_\lambda$, one can define an IP-transition energy interval $ I =[ E^{IP}- \delta, E^{IP}+\delta]$   and look at the partial spectral weight $| \sum_{t\epsilon I}S_{\lambda}^t(\bf{q})| $. This gives information about the absolute contribution to the spectral weight of all transitions of energy close to $E^{IP}$. By choosing many energy intervals, a histogram can be obtained.

However, the IP-weights $S_{\lambda}^t(\bf{q})$ are complex numbers which can sum in constructive or destructive way depending on their phase. In order to account for destructive contributions, the partial spectral weight
\begin{equation}
S_{\lambda}^{E^{IP}}({\bf{q})} = {\bigg| \sum_{t:t\leq {E^{IP}}}S_{\lambda}^t(\bf{q})\bigg|}^2
\end{equation}
is defined as the spectral weight obtained summing only IP-transitions of energy lower or equal to $E^{IP}$. With the help of this quantity, the cumulant weight
\begin{equation}
f_\lambda (E^{IP}) = \lim_{{\bf q}\to {\bf 0}}\frac {\big| S_{\lambda}^{E^{IP}}({\bf{q})}\big|}{\big| S_{\lambda}^{\infty}({\bf{q})}\big|}
\end{equation}
can be plotted as a function of  $E^{IP}$.

\begin{center}
{\bf III. RESULT AND DISCUSSION}
\end{center}

%%%%%%%%%%%%%%%%%%%%%%%%%%%%%%%%%%%%%%%%%%%%%%%%%%%%
\begin{figure}
\centering
\includegraphics[width =\columnwidth] {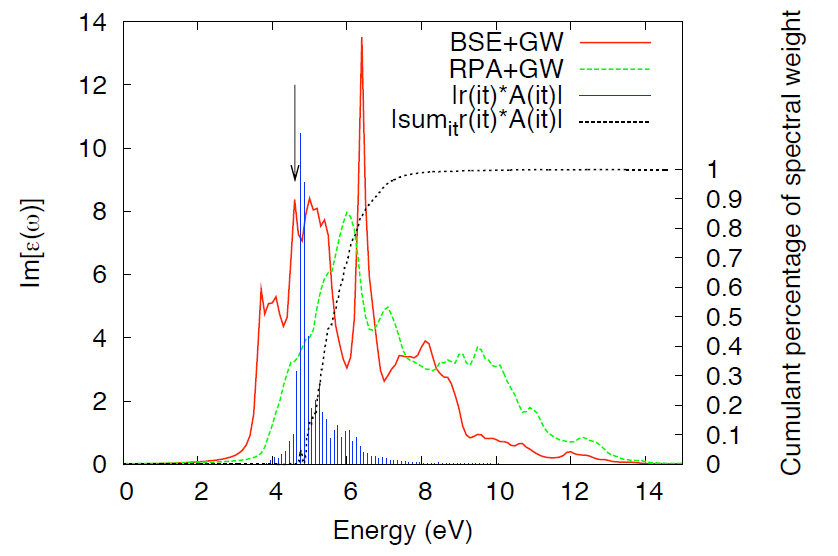}\\[5pt]
\caption{ \label{SupplFig5} (Color online) GW-BSE analysis of the peak Ex$_2$. Peak Ex$_2$: $ E_\lambda= 4.607 $ eV. GW- BSE spectrum and IP-spectrum are represented by the red and green lines respectively. The histogram of $|{\tilde{\rho}}_tA_\lambda^t |$  is presented in blue. The cumulant weight function is reported in dotted black. A black arrow indicates the energy $E_\lambda= 4.607 $eV.}
\end{figure}
%%%%%%%%%%%%%%%%%%%%%%%%%%%%%%%%%%%%%%%%%%%%%%%%%%%%

The temperature dependence reveals a conspicuous evolution of the PDFs. In Fig. \ref{fig1}(a), for the sake of clarity, the PDFs of only the highest (350 K) and lowest (4.2 K) measured temperatures for STO are compared where considerable sharpening of features is observed at the lowest temperature. The most important observations are revealed in the temperature difference data $\delta_T\langle\varepsilon_i\rangle = \langle\varepsilon_i\rangle(T) - \langle\varepsilon_i\rangle(T =350 K), i = 1, 2$ in Figs. \ref{fig1}(b) and \ref{fig1}(c). These results reveal features around 3.35, 3.80, 4.12, 4.70, 4.82, 5.40 and 6.24 eV at 4.2 K (Fig. \ref{fig1}). The structure at 3.35 eV corresponds to the indirect band-gap, which is in agreement with previous reports \cite{Cardona, Benthem, Capizzi}.

%%%%%%%%%%%%%%%%%%%%%%%%%%%%%%%%%%%%%%%%%%%%%%%%%%%%
\begin{figure*}
\centering
\includegraphics[width = 7 in] {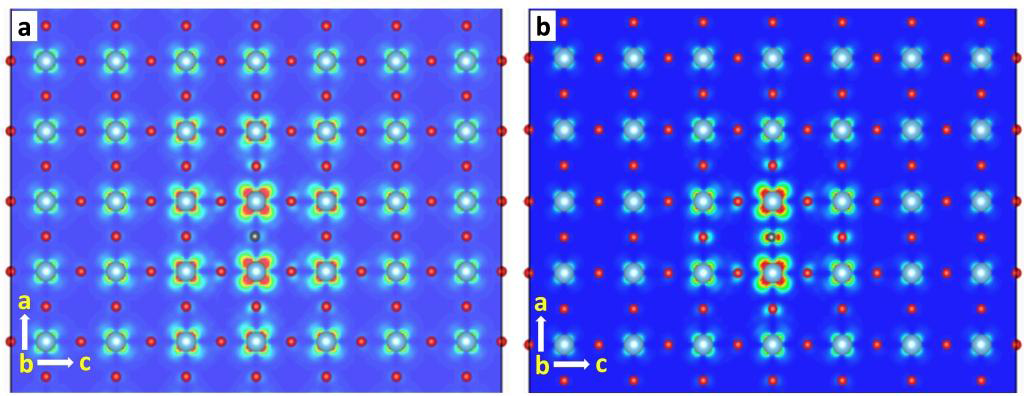}
\caption{\label{fig2} (Color online)
GW-BSE based analysis of excitons Ex$_1$ and Ex$_2$. Excitonic wavefunction $\Psi$(r$_h$, r$_e$) for (a) Ex$_1$ and (b) Ex$_2$ on the \textit{ac} Ti-O plane (red spheres$\colon$oxygen, cyan spheres$\colon$titanium). The probability $\mid$$\Psi$(r$_h$, r$_e$)$\mid$ to find at position  r$_e$ the electron belonging to the exciton when its bound hole (black sphere) is on the central oxygen atom is indicated by the colour scheme from blue to red.}
\end{figure*}
%%%%%%%%%%%%%%%%%%%%%%%%%%%%%%%%%%%%%%%%%%%%%%%%%%%%

The main observation is two sharp peaks at $\sim$3.80 and  $\sim$4.70 eV as clearly seen in $\delta_T\langle\varepsilon_2\rangle$ (Fig. \ref{fig1}(c)). Intriguingly, they both are the most sensitive to temperature, with decreasing temperature their intensities increase monotonically. Furthermore, $\delta_T\langle\varepsilon_1\rangle$ shows a sharp change in phase at these structures (Fig. \ref{fig1}(b)) due to Kramers-Kronig relations between $\varepsilon$$_1$ and $\varepsilon$$_2$. This further supports that these peaks are intrinsic optical properties of STO. Detail fitting based on the second derivative of dielectric function reveals that the these particular peaks are of purely excitonic nature with energy positions at 3.75  and 4.67 eV at 4.2 K (see discussion below). For the rest of discussions, we will refer to these excitons as Ex$_1$ and Ex$_2$, respectively.

We perform two different theoretical calculations, random-phase approximation (RPA) on top of GW (GW-RPA) and GW-BSE calculations, to determine the STO complex dielectric response  $\varepsilon$$_1$ + \textit{i}$\varepsilon$$_2$  (Fig. \ref{fig1}(d)). Note that while the GW-RPA calculations do not account for e-h interactions, the GW-BSE implicitly incorporates e-h and e-e interactions. Intriguingly, the GW-BSE calculations are in remarkable agreement with the experimental data, especially the presence of structures at 3.75 eV and 4.67 eV, which are absent in GW-RPA calculations. This directly implies that the Ex$_1$ and Ex$_2$ features correspond to distinct peaks identified as excitons in the GW-BSE spectrum. Furthermore, comparison of GW-RPA and GW-BSE results reveal a significant spectral weight transfer, a fingerprint of correlations, and hence the crucial role of excitonic effects to explain experimental spectra of STO. More thorough discussions on theoretical calculations are given below.

DFT and G$_0$W$_0$ calculations have been performed with the free plane wave package ABINIT \cite{Gonze}. Norm-conserving pseudopotentials including semicore electrons have been employed for Sr (10 electrons in valence) and Ti (12 electrons in valence). DFT calculations, performed within the local density approximation, converge with a basis set cutoff of 70 Ha. The density has been computed by sampling the Brillouin zone with a 11$\times$11$\times$11 Monkhorst-Pack $k$-point grid, while the Kohn-Sham eigenvalues have been computed on a grid of arbitrarily shifted 512 $k$-points (8$\times$8$\times$8). The G$_0$W$_0$ self-energy has been computed using a cutoff energy of 25 Ha (2400 plane waves) for both the exchange and the correlation terms. The later has been constructed on two frequencies following the Godby-Needs plasmon pole method where 200 bands have been included in the sum over states. The G$_0$W$_0$ corrections have been evaluated on 35 $k$-points (8$\times$8$\times$8 Gamma-centered mesh) and their values have been interpolated to obtain the correction on the shifted grid.

Optical spectra (IP and GW-BSE) have been obtained using G$_0$W$_0$ energies as input. The dielectric matrix has been computed on the 512 shifted $k$-points grid, with a cutoff energy of 40 eV (33 plane waves) and including 40 bands. In GW-BSE calculations, the screened Coulomb interaction has been computed with the same parameters used for the correlation self-energy. Diagonalization of the excitonic Hamiltonian has been performed either in the iterative Haydock scheme with 150 iterations, or in the full diagonalization if also the eigenvalues where needed for analysis.

In the case of SrTi$_{1-x}$Nb$_x$O$_3$, $x =$ 0.005, in order to investigate the evidence of  increase of screening due to doping, we solve GW-BSE using G$_0$W$_0$ energies of the undoped case but accounting for the doping only in the screened Coulomb interaction. This has been computed with the same parameters as the undoped case, but starting from LDA eigenvalues and eigenfunctions obtained including an extra charge of 0.005 electrons per unit cell.

%%%%%%%%%%%%%%%%%%%%%%%%%%%%%%%%%%%%%%%%%%%%%%%%%%%%
\begin{figure}
\centering
\includegraphics[width =\columnwidth] {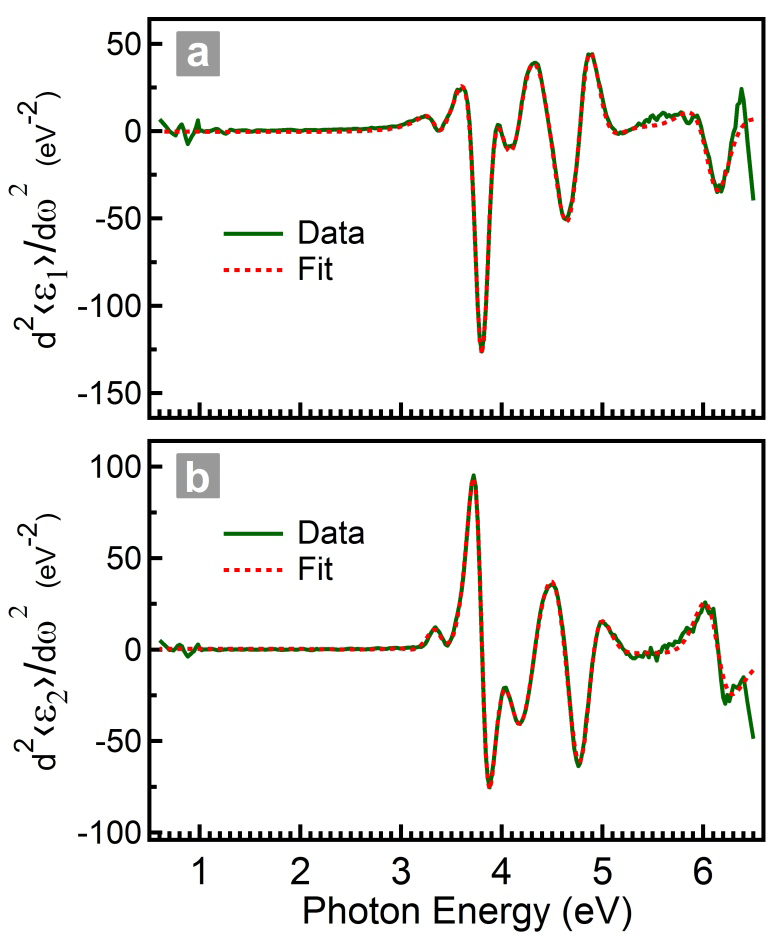}\\[5pt]
\caption{\label{FigDoubleD} (Color online) Double-derivative data and fit of (a) $\langle\varepsilon_1\rangle$ and (b) $\langle\varepsilon_2\rangle$ for STO at 300 K. The fitting parameters can be found in Table \ref{tab1}.}
\end{figure}
%%%%%%%%%%%%%%%%%%%%%%%%%%%%%%%%%%%%%%%%%%%%%%%%%%%%

Figures \ref{SupplFig4} and \ref{SupplFig5} show how the two excitonic peaks (Ex$_1$ and Ex$_2$) are qualitatively different. It has to be noted that  the theoretical $\varepsilon_2$ obtaned using GW-BSE calculations has been plotted with a global blue shift of 47 meV for  comparison with the experimental result in Fig. 1(d) of the main text. Here in the following discussions we refer to the same corresponding peaks but with the associated \textit{theoretical} energy positions. The theoretical peak positions  are 3.644  and 4.607 eV  for Ex$_1$ and Ex$_2$ respectively.  In the case of the peak Ex$_1$  at 3.644  eV (the onset), only IP-transitions of higher energy are summed. This can be clearly seen in Fig. \ref{SupplFig4} by looking at the histogram of $S_t^\lambda$ (blue impulses): it is different from zero only starting from energies $E_\lambda >$ 3.87 eV. This allows for identifying this exciton as a bound exciton since its binding energy is well defined. Moreover, the first contributions are also the most important in absolute value. The analysis shows that the cumulant spectral weight (black dotted line) immediately increases as soon as the histogram has non vanishing contributions. On the other hand it decreases gradually as long as the absolute contributions  become smaller: therefore all  IP-transitions  sum in a constructive way, and all participate to the spectral weight of the first exciton. However, at higher energy (between 6 and 8 eV) the cumulant function goes above 1, having a maximum for 7 eV. This means that if we include IP-transitions up to 7 eV, we will overestimate the intensity of the first peak. Its weight is then reported to the converged result, by including higher energy IP-transitions, which sum with an opposite phase, reducing the spectral weight and taking the cumulant function to the correct asymptotic value of 1. By comparing the cumulant function with the IP-spectrum (green line in Fig. \ref{SupplFig4}), one can see which structures of the IP-spectrum are mixed to create the exciton at 3.644 eV. In this case, contributions are mostly from structures between 4 and 6 eV resulting in a relatively strong excitation effect.

%%%%%%%%%%%%%%%%%%%%%%%%%%%%%%%%%%%%%%%%%%%%%%%%%%%%
\begin{figure}
\includegraphics[width=\columnwidth]{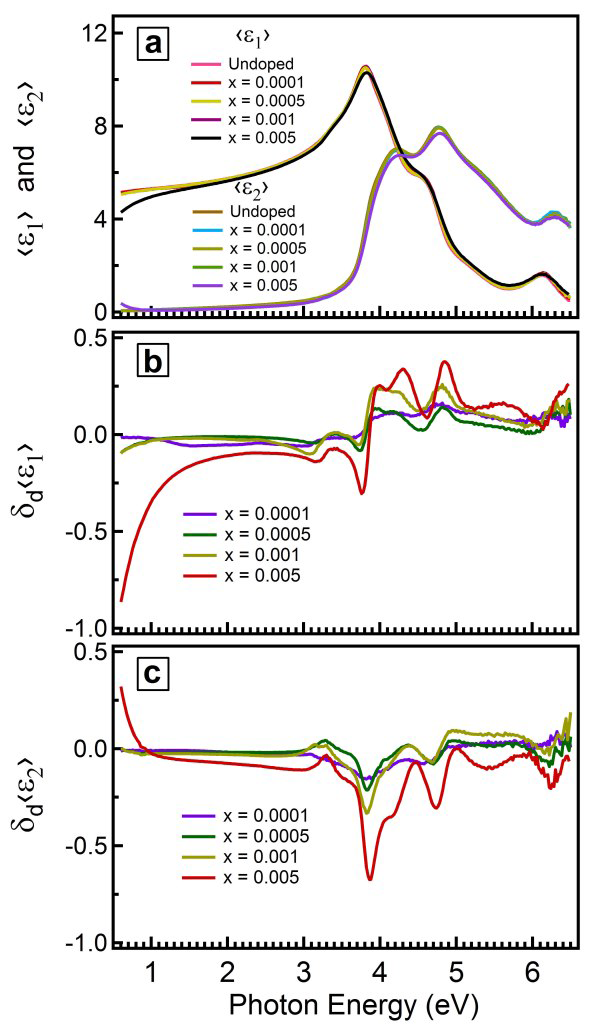}
\caption{\label{fig3} (Color online)
Experimental critical points in the PDFs of STO and SrTi$_{1-x}$Nb$_x$O$_3$. (a) $\langle\varepsilon_1\rangle$  and $\langle\varepsilon_2\rangle$ for STO and SrTi$_{1-x}$Nb$_x$O$_3$ with different doping levels. (b), (c) Doping dependent differences $\delta_d\langle\varepsilon_i\rangle = \langle\varepsilon_i\rangle(x) - \langle\varepsilon_i\rangle(x = 0), i = 1, 2$.}
\end{figure}
%%%%%%%%%%%%%%%%%%%%%%%%%%%%%%%%%%%%%%%%%%%%%%%%%%%%

\begin{center}
\begin{table*}
{\caption{Parameters of the double derivative fit of $\langle\varepsilon\rangle$ for STO at 300 K.} \label{tab1}
\renewcommand{\arraystretch}{1.2}
\begin{tabular}{|c|c|c|c|c|c|}
\hline
 \textbf{Critical Points} & \textbf{E (eV)} & \textbf{A} & \textbf{$\Gamma$ (meV)} & \textbf{$\phi$ (deg)} & \textbf{Type}\\

\hline
E$_0$ &	3.350(29) &	0.27(1)	& 156(31) &	303.62(38) &	2D\\
\hline
Ex$_1$ &	3.801(4) &	0.43(3) &	193(4) &	86.45(6) &	Excitonic\\
\hline
E$_1$	 & 4.147(16)	 & 4.63(85) &	277(19) &	311.96(15) &	2D\\
\hline
Ex$_2$ &	4.712(19) &	8.40(2.34)	 &  531(21) &	174.24(5) &	Excitonic\\
\hline
E$_2$  & 	4.860(1) &	8.39(2.95)	 & 264(20) &	195.06(3)	& 2D\\
\hline
E$_3$  &	6.000(338)	& 16.63(6.76)   & 1537(356) &  73.49(31)	& 2D\\
\hline
Ex$_3$ &	6.113(10)	& 0.58(8) &	325(14) &	255.62(10)	& Excitonic\\
\hline

\end{tabular}}
\end{table*}
\end{center}

The analysis of the peak Ex$_2$ leads to a different interpretation. As it can be seen in Fig. \ref{SupplFig5} by the histogram  $S_t^\lambda$ (in blue), IP-transitions of energy $E_\lambda <$ 4.644 eV are included in the summation, which makes it difficult to give a clear-cut definition of its \textit{binding energy}. However, the fact that the cumulant weight function (dotted black) is vanishing, tells us that these transitions have opposite phase and they sum destructively. It is interesting to note also that the most important absolute contributions (that is the highest blue impulses) have opposite phases. This is evident from the small peak in the cumulant function: it raises because of the first impulse, but it is immediately killed by the transitions of slightly higher energy (the second impulse) which have opposite phase. Amongst the transitions summing constructively in the peak Ex$_2$, the bump at 7 eV (green line) contributes for $\sim$20\%.

Based on GW-BSE calculations, the origin of the various structures in the optical spectra of STO can now be interpreted more comprehensively. Analysis of the exciton composition in terms of the remixing of single-particle e-h transitions confirms the strong excitonic nature of both peaks (see above). The dominant contributions to Ex$_1$  are coming from O-\textit{2p}$\to$ Ti-\textit{3d} t$_{2g}$ transitions in the range from the band-gap to $\sim$4.5 eV. Thus, Ex$_1$ can clearly be identified as a bound exciton. The calculated binding energy for Ex$_1$ is $\sim$220 meV, which is in good agreement with our experimental estimation. In contrast, even though Ex$_2$ shares the same O-\textit{2p}$\to$ Ti-\textit{3d} t$_{2g}$ orbital character as Ex$_1$, the dominant contributions are from larger energies (from $\sim$ 4.5 to $\sim$ 5 eV).  Therefore, Ex$_2$ can be interpreted as a resonant exciton.

The real space plot of the excitonic wavefunction $\mid$$\Psi$(r$_h$, r$_e$)$\mid$ as calculated by BSE for the bound exciton Ex$_1$ is shown in Fig. \ref{fig2}(a). The shape of the exciton wavefunction clearly has the characteristic of the Ti-\textit{3d} t$_{2g}$ electrons, consistent with the above-mentioned compositional analysis. However, surprisingly the bound exciton Ex$_1$ appears as a highly delocalized Wannier-\textit{like} exciton emerging from strongly localized Ti-\textit{3d}  electrons. Meanwhile, the resonant Ex$_2$ exhibits a more localized character than the bound Ex$_1$, which is contrary to the conventional Wannier picture \cite{YuCardona}, as seen in Fig. \ref{fig2}(b).

Finally, a feature labelled Ex$_3$ is observed at 6.11 eV (this energy position is determined using double-derivative analysis as shown below. Based on BSE calculations at higher energy bands \cite{Sponza}, this feature has been attributed to resonant exciton via O-\textit{2p}$\to$ Ti-\textit{3d} e$_{g}$  transitions with spectral weight mostly from states at higher energy of 8-10 eV. In order to analyze the detailed temperature dependence, it requires optical spectra with much higher energy, which is beyond our current measurement. Nevertheless, this further supports the importance of electronic correlations in STO. Furthermore, the prominent features observed experimentally at 4.12 eV, 4.82 eV, 5.40 eV are inter-band transitions as shown in our theoretical results.

To further support our analysis, we perform double-derivative analysis of critical points of STO dielectric function at 300 K as representatives. The structures in the optical spectra ($\varepsilon$) can be resolved into multiple critical points using standard analytical line-shapes \cite{MCardona, Daspnes, Lautenschlager, Zollner} given by

\begin{equation} \label{eqn9}
\varepsilon(\omega) = C - Ae^{i\phi}{(\omega - E + i\Gamma})^n
\end{equation}

Here the critical point parameters $A$, $E$, $\Gamma$ and $\phi$  represent amplitude, threshold energy, broadening and excitonic phase angle respectively. The exponent $n$ depends on the dimension of the critical point in consideration.

Double derivative fit of the (pseudo-) dielectric function of STO and SrTi$_{1-x}$Nb$_x$O$_3$, 0$\leq$\textit{x}$\leq$0.005 have been performed for all the measured temperatures. The equations used for fitting are  derived from Equation \ref{eqn9} and are given below.

For a discrete excitonic critical point  lineshape

\begin{equation}
\frac{d^2{\langle\varepsilon\rangle}}{d{\omega}^2} = n(n-1)Ae^{i\phi}{(\omega - E + i\Gamma})^{n-2}, n=-1
\end{equation}

and for a two dimensional (2D) critical point lineshape

\begin{equation}
\frac{d^2{\langle\varepsilon\rangle}}{d{\omega}^2} = Ae^{i\phi}{(\omega - E + i\Gamma})^{-2}
\end{equation}

A representative example of such fitting analysis for the critical points of STO at 300 K is presented here in Fig. \ref{FigDoubleD}.  The fitting parameters for the best fit (with MSE of  4.287) are given in Table \ref{tab1}. The numbers in the parentheses indicate 95\% confidence limit.

%%%%%%%%%%%%%%%%%%%%%%%%%%%%%%%%%%%%%%%%%%%%%%%%%%%%
\begin{figure}
\includegraphics[width=\columnwidth]{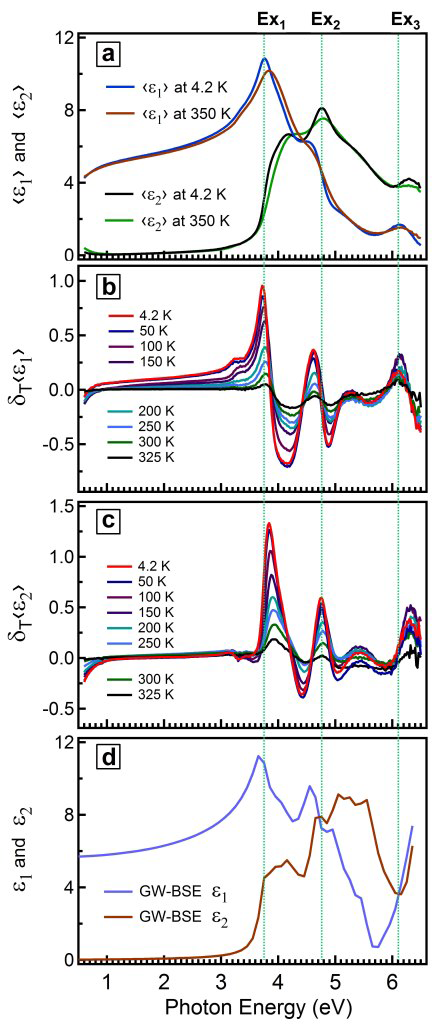}
\caption{\label{fig4} (Color online)
Experimental critical points and comparison with theory for SrTi$_{1-x}$Nb$_x$O$_3$, \textit{x} = 0.005. (a) $\langle\varepsilon_1\rangle$  and $\langle\varepsilon_2\rangle$  for the lowest (4.2 K) and highest (350 K) measured temperatures. (b), (c) Temperature dependent differences $\delta_T\langle\varepsilon_i\rangle = \langle\varepsilon_i\rangle(T) - \langle\varepsilon_i\rangle(T =350 K), i = 1, 2$.  (d)  Plots of $\varepsilon$$_1$ and $\varepsilon$$_2$ calculated using GW-BSE for SrTi$_{1-x}$Nb$_x$O$_3$, \textit{x} = 0.005 (all graphs are blue-shifted by 53 meV to match experimental data). The green vertical dashed line are determined using second derivative analysis.}
\end{figure}
%%%%%%%%%%%%%%%%%%%%%%%%%%%%%%%%%%%%%%%%%%%%%%%%%%%%

The evolution of the optical spectra of SrTi$_{1-x}$Nb$_x$O$_3$ with different  $x$ (0.0001, 0.0005, 0.001 and 0.005) is shown in Fig. \ref{fig3}.  The room temperature plots of $\langle\varepsilon_1\rangle$  and $\langle\varepsilon_2\rangle$ (Fig. \ref{fig3}(a)) exhibit a strong  Drude tail for the sample with highest doping ($x = 0.005$). Differences $\delta_d\langle\varepsilon_1\rangle$ and $\delta_d\langle\varepsilon_2\rangle$ between the PDF of STO and SrTi$_{1-x}$Nb$_x$O$_3$ ($\delta_d\langle\varepsilon_i\rangle = \langle\varepsilon_i\rangle(x) - \langle\varepsilon_i\rangle(x = 0), i = 1, 2$) highlight the doping-dependent evolutions as shown in Figs. \ref{fig3}(b) and \ref{fig3}(c), respectively. The changes are manifested as a Drude tail and several negative peaks in $\delta_d\langle\varepsilon_2\rangle$. The peak positions gradually shift to higher energies with increasing doping.

Another important observation is the evolution of the excitonic peaks and the corresponding rearrangement of the spectral weight with the variation of $x$ (Fig. \ref{fig3}(c)). With the increase of $x$ from 0 to 0.001 it is seen that the spectral weight is redistributed only between the energy range from $\sim$3 to 6.5 eV. For example, the negative peak at $\sim$3.9 eV is compensated by the positive contributions from above $\sim$4.8 eV. However, in the case of \textit{x} = 0.005 the onset of the Drude tail is seen along with a change in trend of the spectral weight transfer. Spectral weight redistribution is not confined to the energy range from $\sim$3 to 6.5 eV anymore as before, conspicuously changing to negative above $\sim$ 4.8 eV. This could be an indication that SrTi$_{1-x}$Nb$_x$O$_3$  is still in the insulating state for $x\leq0.001$ despite the presence of doping.

In Figure \ref{fig4}(a) the PDFs at 4.2 K and 350 K are plotted for $x =0.005$. The PDF temperature difference plots in Figs. \ref{fig4}(b) and \ref{fig4}(c) ($\delta_T\langle\varepsilon_1\rangle$ and $\delta_T\langle\varepsilon_2\rangle$, respectively) show signatures of a Drude tail and sharper indirect band-edge features. The enhancement of the indirect band-edge feature could be due to stronger electron-phonon coupling \cite{vanMechelen}.  Similarly to the case of STO, excitonic peaks are seen as prominent features in $\delta_T\langle\varepsilon_2\rangle$. From double derivative analysis these peaks are found to be at 3.75, 4.76 and 6.11 eV as compared to 3.75, 4.67  and 6.11 eV, respectively, in case of STO. The blue-shift in the case of the localized exciton Ex$_2$, together with broadening and decrease of heights of all the peaks, can be interpreted as a reduction of the excitonic effect due to the presence of Nb-dopant free charge carriers which increases the screening and reduces the e-h interaction. Systematic measurements performed also on SrTi$_{1-x}$Nb$_x$O$_3$, $x= 0.0005$ (which is an intermediate value) show similar trend. This mechanism is confirmed by GW-BSE simulations (Fig. \ref{fig4}(d)) with doping and these results represent further evidence of the dominant excitonic nature of these peaks. The dielectric function calculated from GW-BSE for SrTi$_{1-x}$Nb$_x$O$_3$, $x = 0.005$ are plotted in Fig. \ref{fig4}(d). There is very good agreement with the experimental PDF results in terms of excitonic peak positions as well as other prominent features.

\begin{center}
{\bf IV. CONCLUSION}
\end{center}

In summary, the evidence of anomalous excitonic effects in the optical spectra of STO in the form of a Wannier-\textit{like} bound exciton at 3.75 eV, resonant excitons at 4.67 eV and 6.11 eV at 4.2 K, as well as their evolution with temperature, doping-modulated screening and correlation, are reported. Remarkably the Wannier-\textit{like} bound exciton is more delocalized than the resonant exciton. Theoretical calculations further show the indispensable role e-h and e-e interactions, yielding to excitonic effects, play in understanding the dielectric function as a whole. These observations of prominent and novel correlated phenomena in STO may pave the way for correct understanding of not only STO \textit{per se} but also material systems where it is an integral component such as heterostructures.

\begin{center}
{\bf V. ACKNOWLEDGMENT}
\end{center}

We acknowledge valuable discussions with T. Venkatesan, Giovanni Vignale, Ariando and Michael R\"{u}bhausen.  This work is supported by Singapore National Research Foundation under its Competitive Research Funding (NRF-CRP 8-2011-06 and NRF2008NRF-CRP002024), MOE-AcRF Tier-2 (MOE2010-T2-2-121), NUS-YIA. We acknowledge the CSE-NUS computing centre and GENCI for providing facilities for our numerical calculations. PTE also acknowledges the National Research Foundation, Prime Minister�s Office, Singapore, under its Medium Sized Centre Programme and Competitive Research Funding (R-144-000-295-281). A part of this work was supported by the Use-of-UVSOR Facility Program (BL3B, Proposal No. 26-561) of the Institute for Molecular Science.

% Bibliography
%

%\bibliographystyle{apsrev4-1}
%\bibliography{bibliographyn}


\end{document}